
\input harvmac
%
%
%
%
\ifx\answ\bigans
\else
\output={
  \almostshipout{\leftline{\vbox{\pagebody\makefootline}}}
         \advancepageno
}
\fi
%
%
%
\def\mayer{\vbox{\baselineskip=14truept
\sl\centerline{Department of Physics 0319}%
\centerline{University of California, San Diego}
\centerline{9500 Gilman Drive}
\centerline{La Jolla, CA 92093-0319}}}
%
%

%
%
%
%
\def\abstract#1{\centerline{\bf Abstract}\nobreak\medskip\nobreak\par

#1}
%
%
%
%
%



%
%
%
%
\def\inv{^{\raise.15ex\hbox{${\scriptscriptstyle -}$}\kern-.05em 1}}
\def\lbar{{\lower.35ex\hbox{$\mathchar'26$}\mkern-10mu\lambda}}


%
%
%
%
\def\dsl{\,\raise.15ex\hbox{/}\mkern-13.5mu D}

\def\delsl{\raise.15ex\hbox{/}\kern-.57em\partial}
\def\Ksl{\hbox{/\kern-.6000em\rm K}}
\def\Asl{\hbox{/\kern-.6500em \rm A}}
\def\Dsl{\hbox{/\kern-.6000em\rm D}} 
\def\Qsl{\hbox{/\kern-.6000em\rm Q}}
\def\gradsl{\hbox{/\kern-.6500em$\nabla$}}
%
%
\def\lspace{\ifx\answ\bigans{}\else\qquad\fi}
\def\lbspace{\ifx\answ\bigans{}\else\hskip-.2in\fi} 
%
%
\def\boxeqn#1{\vcenter{\vbox{\hrule\hbox{\vrule\kern3pt\vbox{\kern3pt
        \hbox{${\displaystyle #1}$}\kern3pt}\kern3pt\vrule}\hrule}}}
%
%
\def\mbox#1#2{\vcenter{\hrule \hbox{\vrule height#2in
\kern#1in \vrule} \hrule}}
%
%
%
%

\def\CC{{\cal C}}

 \def\CH{{\cal H}}

%
%
%
%
%

%

\def\bar#1{\overline{#1}}

\def\darr#1{\raise1.5ex\hbox{$\leftrightarrow$}\mkern-16.5mu #1}

%
%

\def\frac#1#2{{\textstyle{#1\over #2}}} 
%
%
%
%

\def\Im{\mathop{\rm Im}}
\def\Re{\mathop{\rm Re}}

%
%
%
%

%
%
\def\ltap{\ \raise.3ex\hbox{$<$\kern-.75em\lower1ex\hbox{$\sim$}}\ }
\def\gtap{\ \raise.3ex\hbox{$>$\kern-.75em\lower1ex\hbox{$\sim$}}\ }
\def\gl{\ \raise.5ex\hbox{$>$}\kern-.8em\lower.5ex\hbox{$<$}\ }
\def\roughly#1{\raise.3ex\hbox{$#1$\kern-.75em\lower1ex\hbox{$\sim$}}
}
%
%

%

%

\relax

\def\dpart{\partial\kern .5ex\llap{\raise
1.7ex\hbox{$\leftrightarrow$}}\kern -.7ex {_\mu}}

\def\frac#1#2{{\textstyle{#1 \over #2}}}

\def\Im{{\rm Im}}
\def\Re{{\rm Re}}

\def\s2weak{\sin^2\theta_{\rm w}}

\def\({\left(}\def\){\right)}

\def\mayer{\vbox{\baselineskip=14truept\sl\centerline{Department of
Physics,
9500 Gilman Drive 0319}
\centerline{University of California, San Diego}
\centerline{La Jolla, CA 92093-0319}}}

\def\Queens{\vbox{\baselineskip=14truept
\sl\centerline{Department of Physics, Stirling Hall}
\centerline{Queen's University}
\centerline{Kingston, Canada, K7L 3N6}}}
\def\Duke{\vbox{\baselineskip=14truept
\sl\centerline{Department of Physics}
\centerline{Duke University, Durham, NC 27708-0305}}}

\def\[{\left[}
\def\]{\right]}
\def\({\left(}
\def\){\right)}

\noblackbox
 at 12truept
\vskip 1.in
\centerline{{\titlefont{E2/M1 Mixing Ratio of $\Delta\rightarrow
N\gamma$ and }}}
\medskip
\centerline{{\titlefont{Hyperon Resonance Radiative Decay}}}
\bigskip
\centerline{Malcolm N.\ Butler}
\bigskip
\Queens
\bigskip
\centerline{Martin J.\ Savage\footnote{$^{\dagger}$}
{SSC Fellow}}
\bigskip
\mayer
\bigskip
\centerline{Roxanne P.\ Springer}
\bigskip
\Duke
\bigskip
\centerline{{\bf Abstract}}
\medskip

We compute the leading contribution to the E2/M1 mixing ratio of the
decay $\Delta\rightarrow N\gamma$ in heavy baryon chiral perturbation
theory.  We find the mixing ratio to be $4\% \ltap |\delta_{\rm E2/M1}|
\ltap 11\%$, much larger than estimates based on the quark model and
other hadronic models.  We also compute the mixing ratio for the
radiative decay of the hyperon  resonances.   The decays
$\Sigma^{*+}\rightarrow\Sigma^+\gamma$ and
$\Xi^{*-}\rightarrow\Xi^-\gamma$ provide a particularly sensitive probe
of deviations from heavy baryon spin-flavour SU(6).

\vfill
\line{UCSD/PTH 93-04, QUSTH-93-01, Duke-TH-93-47 \hfil January 1993}
\line{hep-ph/9302214 \hfil}
\eject

The ratio of electric quadrupole radiation (E2) to magnetic dipole
radiation (M1) in the decay $\Delta\rightarrow N\gamma$ is of great
interest as it provides valuable information on the ground state
structure of  the lowest lying baryons.  In the simplest constituent
quark model this radiative transition is purely M1 arising from the
spin flip of a single constituent quark.  However, the hyperfine
interaction between constituent quarks induces admixtures of non-zero
orbital angular momentum states into the hadronic ground states.  Such
configurations give rise to non-vanishing E2 matrix elements.  We will
denote the ratio of reduced matrix elements for E2 radiation to M1
radiation by $\delta_{\rm E2/M1}$.  Pion--photoproduction has been used
to arrive at an experimental determination of this mixing ratio, but
only after model dependent analyses have been performed \ref\photoa{H.
Tanabe and K. Ohta, Phys. Rev. {\bf C31} (1985) 1876.}\nref\photob{ R.
M. Davidson, N.C. Mukhopadhyay and R.S. Wittman,  Phys. Rev. {\bf D43}
(1991) 71.}\nref\photoc{ R.L. Workman, R. A. Arndt and Z. Li, Virginia
Polytechnic Institute and State University Preprint}-\ref\photod{ S.
Nozawa, B. Blankleider and T.S.H. Lee,  Nuc. Phys. {\bf A513} (1990)
459.}.  These analyses would indicate that $-3\% <  \delta_{\rm E2/M1}
< +4\%$, where only local counterterms have been used in forming
$\delta_{\rm E2/M1}$, forcing it to be real and model dependent.
Recent lattice computations \ref\latta{D. Leinweber, T. Draper and R.M.
Woloshyn, Univ. of Maryland Preprint, U.MD PP \#93-085.} constrain the
mixing ratio to lie in the interval $-5\% <  \delta_{\rm E2/M1} <
11\%$. Since lattice calculations treat the $\Delta$ as a stable
particle, this ratio is also real.  The nonrelativistic constituent
quark model predicts $\delta_{\rm E2/M1}$ to be very small on the order
of $-0.4$\%, and to be essentially zero when relativistic wavefunctions
are used~\ref\capstick{S.\ Capstick, Phys.\ Rev. \ {\bf D46} (1992)
1965; {\bf D46} (1992) 2864.}.  However, it has been suggested that
contributions from the pion cloud are likely to dominate this mixing
ratio~\ref\pionic{S.\ Kumano, Phys.\ Lett.\ {\bf B214} (1988) 132}.  In
this work we will compute the leading, model independent contribution
to $\delta_{\rm E2/M1}$ in heavy baryon chiral perturbation theory
(HBChPT).

Unlike the M1 amplitude, which is dominated by short distance physics
through a local counterterm and kaon loops, the E2 amplitude is
dominated by long distance pion loops which are enhanced by factors of
$\log(m_\pi^2/\Lambda_\chi^2)$ over naive estimates for the size of the
E2 counterterm.  Such chiral logs can be unambiguously computed in
HBChPT, and as these are formally the leading contribution to the E2
amplitude we will use them to predict the mixing ratios $\delta_{\rm
E2/M1}$.  We will not review here the formalism for chiral perturbation
theory when the baryons are treated as heavy, the reader should refer
to \ref\heavybaryon{E.\ Jenkins and A.\ Manohar, Phys.\ Lett.\ {\bf
255B} (1991) 558 ; Phys.\ Lett.\ {\bf 259B} (1991) 353.} (for a review
see \ref\jmhungary{E.  Jenkins and A. Manohar, Proceedings of the
workshop on``Effective Field Theories of the Standard Model,'' ed.  U.
Meissner, World Scientific (1992)}).

In an earlier work \ref\butlera{M.N. Butler, M.J. Savage and R.P.
Springer, UCSD/PTH 92-37,QUSTH-92-05, Duke-TH-92-44}, we studied the
electromagnetic branching fractions of the decuplet of baryons using
HBChPT, and also the strong interaction couplings at one--loop.  One
interesting outcome was the striking similarity in the relative
strengths of the strong couplings extracted from measured widths,
and the relationships predicted by heavy baryon spin--flavour
SU(6)  \jmhungary\ref\carone{C. Carone and H. Georgi, Nuc. Phys.
{\bf B375} (1992) 243.}.  This would seems to indicate some
manifestation of a higher symmetry group than the SU(3) of QCD.  To
investigate this further, we can compare the HBChPT predictions for
$\delta_{\rm E2/M1}$, using the independent values of the strong
couplings from SU(3), to predictions when the SU(6) relations between
the couplings are imposed.

There is one dimension five counterterm that contributes
to the radiative decay of the baryon decuplet of the form
\eqn\trans{{\cal L}^{\rm M1}_v = i \Theta_1{e\over \Lambda_\chi} \bar
B_v
S^\mu_v Q T_v^\nu F_{\mu\nu}\ \ \ \ \ ,} where $Q$ is the
electromagnetic
charge matrix,
\eqn\charge{Q = \left(\matrix{ 2/3&0&0\cr 0&-1/3&0\cr
0&0&-1/3}\right) \ \ \
\ ,}
(to lowest order in meson fields), and $\Theta_1$ is an unknown
coefficient that was determined in \butlera\  by fitting to the
branching ratio for $\Delta\rightarrow N\gamma$.  This local
counterterm contributes only to the M1 transition as can easily be seen
by going to the baryon rest frame, where $\mu,\nu$ are spacelike only.
The $\pi$ and $K$ loops contribute at order $1/\Lambda_\chi^2$.
Consequently, the rates for the SU(3) allowed transitions are dominated
by the local counter term while the SU(3) forbidden transitions are
dominated by the loops.  There is a dimension six counterterm that
contributes to the  E2 amplitude of the form
\eqn\trans{{\cal L}^{\rm E2}_v = i \Theta_2{e\over \Lambda_\chi^2}
\bar B_v
(S^\mu_v Q T_v^\nu + S^\nu_v Q T_v^\mu )v^\alpha\partial_\mu
F_{\alpha\nu}\ \ \ \ \ ,}
whose coefficient $\Theta_2$ is also unknown.   However,  as mentioned
earlier,
this local counterterm is formally
subleading to the contribution from the $\pi$ loops  which
are enhanced by a factor of
$\log(m_\pi^2/\Lambda_\chi^2)$ over the naive estimate of $\Theta_2$,
so  we will neglect $\Theta_2$ for the purposes of this work.
The matrix element for radiative decays can be written as \butlera
\eqn\matel{{\cal M} = X\ \ \overline{B}S\cdot k\
T\cdot\epsilon_\gamma\ \ \
+\ \ \ Y\ \ \overline{B} S\cdot\epsilon_\gamma\ k\cdot T\ \ \ \ ,}
where $X$
and $Y$ include contributions from both one-loop graphs and the M1 local
counterterm.
The coefficients $X$ and $Y$,  which were computed in \butlera,
are functions of the mass of the $\pi$ or $K$, the mass splitting
between the initial and
intermediate state baryon, and the energy of the emitted photon.
They are given by
\eqn\xx{\eqalign{
X =\ & \ -ie\left[ Q_{TB}{\Theta_1\over \Lambda_\chi} +
{1\over 4\pi^2f^2} \bigg(\beta_{TBB}I_2-\beta_{TTB}[I_1-{2\over
3}I_2]\bigg)\right]\cr
 Y =\ & \ -ie\left[ -Q_{TB}{\Theta_1\over\Lambda_\chi}+
{1\over 4\pi^2f^2} \bigg(\beta_{TBB}I_1-\beta_{TTB}[I_2-{2\over
3}I_1]\bigg)\right]\cr}\ \ \ \ ,}
where $Q_{TB}$, $\beta_{TBB}$ and $\beta_{TTB}$ are Clebsch--Gordan
coefficients
that can be found
in \butlera,  and $f$ is the meson decay constant; $f_\pi$ for
$\pi$'s and $f_K$ for $K$'s.
The integrals $I_1$ and $I_2$ are given by
\eqn\integs{\eqalign{I_1&=\int_0^1dx x
\Gamma(\epsilon) I(-\epsilon,\omega_\gamma x-\Delta m,M^2)\cr
I_2&=\int_0^1dx(x-1)\Gamma(\epsilon)I(-\epsilon,\omega_\gamma
x-\Delta m,
M^2)\cr}\ \ \ \ \ ,}
where $\omega_\gamma$ is the energy of the photon emitted during the
decay, $\Delta m$ is the mass difference between the initial decuplet
baryon and the intermediate state baryon, and $M$ is the mass of either
the $\pi$ or $K$ in the loop.  An SU(3) symmetric $\overline{\rm MS}$
subtraction scheme is used to define the finite parts of $I_1$ and
$I_2$, and  we use
\eqn\subtract{\Gamma(\epsilon)I(-\epsilon,b,c) = b\left[
\log(c/\Lambda_\chi^2)
-2\right]-\sqrt{b^2-c}\log\left({b-\sqrt{b^2-c+i\epsilon}\over
b+\sqrt{b^2-c+i\epsilon}}\right)\ \ \ \ ,} where only finite pieces
are shown.

The E2 and M1 amplitudes can be found from the symmetric and
antisymmetric combinations of
$X$ and $Y$ respectively,
\eqn\multi{\eqalign{A_{\rm E2} &= {\sqrt{5}\over
2}\left(X+Y\right)\cr
A_{\rm M1} &= {1\over
2}\left(X-Y\right)}\ \ \ \ ,}
from which we find that the total width for the decay is
\eqn\rate{\Gamma (T\rightarrow B\gamma ) = {\omega_\gamma^3\over
12\pi}\left[ |A_{\rm E2}|^2
+ |A_{\rm M1}|^2 \right].}
We note that the overall sign of $A_{\rm E2}$ cannot be determined
uniquely, and that we have chosen the sign defined in Eq.\ \multi.  The
mixing ratio, $\delta_{\rm E2/M1}$, is defined as
\eqn\mix{\delta_{\rm E2/M1} = {1\over\sqrt{3}}{A_{\rm E2}\over A_{\rm
M1}}\ \ \ \ \ ,}
using the common convention for this quantity.
The rates computed in \butlera\  using \rate\  agree, within errors,
with the lattice computations of \latta .
However, these rates are dominated by the $M1$ amplitude, where
a true test of these methods is in the E2 component.

The  M1 counterterm is determined by the rate for $\Delta\rightarrow
N\gamma$ and consequently depends on the choice of strong couplings
constants $F,\ D$, $\CC$ and  $\CH$.  These couplings and their
associated uncertainties have been found from axial current matrix
elements amongst the octet baryons\heavybaryon\ , from the mass
splitting of the lowest lying baryons and baryon resonances
\ref\liza{E. Jenkins, Nuc. Phys. {\bf B368} (1992) 190.}, from
nonleptonic hyperon  decays \ref\lizb{E. Jenkins, Nuc. Phys. {\bf B375}
(1992) 561.}, and from the strong decay of baryon resonances \butlera.
In order to find the predicted value for each mixing ratio we randomly
chose values for the constants $F,\ D$, $\CC$ and $\CH$ from the
intervals $0.5 < D < 0.7$, $0.3 < F < 0.5$, $-1.3 < \CC < -1.1$ and
$-2.8 < \CH < -1.6$.  As there is an uncertainty in the radiative width
for $\Delta\rightarrow N\gamma$, we also randomly chose a value for the
radiative width from the allowed $2\sigma$ region \ref\pdg{Particle
Data Group, Phys. Rev {\bf D45} (1992) 1.}.  With these values we
solved for the M1 counterterm $\Theta_1$, and then computed the mixing
ratio $\delta_{\rm E2/M1}$.  We studied a sample of 6000 points in the
allowed region, and the results are shown in \fig\allow{The E2/M1
ratio  ($\delta = \delta_{\rm E2/M1}$) for the SU(3) allowed radiative
decays of the decuplet.  The gray region corresponds to the 6000 points
randomly chosen only from the uncertainties in the couplings
$F,\ D,\ \CC$, $\CH$, and the width for $\Delta\to N\gamma$.  The dark
lines are the predictions of heavy baryon SU(6), chosen from the
uncertainty in $D$, and still including the uncertainty in the
$\Delta\to N\gamma$ width.  We have assumed the E2 counterterm to be
small compared to the contribution from pion loops.} and \fig\supp{The
E2/M1 ratio  ($\delta = \delta_{\rm E2/M1}$) for the SU(3) suppressed
radiative decays of the decuplet.  The gray region corresponds to the
6000 points randomly chosen from the uncertainties in the couplings $F,
D, \CC$, and $\CH$.  The dark points are the predictions of heavy baryon
SU(6).  Note that these SU(6) predictions are independent of the choice
for $D$, and independent of the $\Delta\to N\gamma$ decay width.
(These decays do not receive a contribution from either local
counterterm $\Theta_1$ or $\Theta_2$ because of the lowest order U-spin
invariance of electromagnetic interactions.)} as the gray region.

It is interesting to see what the constraints imposed by a heavy baryon
spin-flavour SU(6) \jmhungary--\carone  would have on our
predictions.  Such a symmetry relates all the lowest order strong
coupling constants, $F={2\over 3}D$, $\CC=-2D$ and $\CH=-3D$. We chose
a value of $D$ from the interval $0.5 < D < 0.7$ and then predicted
each mixing ratio, the results of which appear as the dark lines in
\allow\ and dark points in \supp.  Note that the SU(3) forbidden decays
are independent of the choice of the coupling constant $D$, since it
divides out in the ratio, yielding sharp predictions based on heavy
baryon spin--flavour SU(6) for these transitions, depending only on
baryon masses, meson masses, and photon energy.

Typical values of the ratio for the SU(3) forbidden decays
($\Sigma^{*-}\to\Sigma^-\gamma$ and $\Xi^{*-}\to\Xi^-\gamma$) are large
and might be measured at CEBAF.  It has also been suggested that
polarized hyperon beams at FNAL could be used~\ref\fnal{H.J.\ Lipkin
and M.A.\ Moinester, Phys.\ Lett.\ {\bf B287} (1992) 179.}.  Again,
these SU(3) suppressed  decay modes  are important in that they are
dominated by loops and do not depend on the counterterms $\Theta_1$ and
$\Theta_2$ ( the same U--spin argument that forbids a contribution from
$\Theta_1$ also forbids a contribution from $\Theta_2$).  We do not
show $\delta_{\rm E2/M1}$ for the decay $\Sigma^{*0}\to\Sigma^0\gamma$.
Since the one-loop contribution to its  imaginary part vanishes in the
limit of exact isospin symmetry (from exact cancellations between
contributing intermediate states),
 this mixing ratio is very sensitive to isospin breaking and  reliable
predictions for this particular mode cannot be made at this order.

An important observation can be made from \allow\ and \supp.  The
decays $\Sigma^{*+}\rightarrow\Sigma^+\gamma$ and
$\Xi^{*-}\rightarrow\Xi^-\gamma$ are particularly sensitive to
deviations from heavy baryon spin-flavour SU(6).  This symmetry
predicts that $\Re(\delta_{\rm E2/M1})$ for the transition
$\Sigma^{*+}\rightarrow\Sigma^+\gamma$  is very much smaller than
$\Im(\delta_{\rm E2/M1})$ while the possible range of values determined
from the experimental uncertainties in the couplings constants {\sl
without} imposing the  SU(6) relations is large.  A similar tendency is
seen for the $\Sigma^{*-}\rightarrow\Sigma^-\gamma$ result.  For both
this transition and for $\Xi^{*-}\rightarrow\Xi^-\gamma$, the SU(6)
prediction of $\delta_{\rm E2/M1}$ is very restrictive, and at values
large enough to be more experimentally accessible than some of the
regions allowed without imposing this symmetry.

We should comment on the presence of an imaginary part in $\delta_{\rm
E2/M1}$. This arises from final--state interactions where the
intermediate state is an octet baryon and a pion; a strong decay mode
of the decuplet baryon.  As mentioned, most pion--photoproduction
analyses attempt to extract the contribution to $\delta_{\rm E2/M1}$
arising from local counterterms only\photoa--\photod.  Such an
extraction will necessarily yield a $\delta_{\rm E2/M1}$ which is real,
and relegates the pionic dressing of the vertex to be a background
contribution.  A comparison of the value of $\delta_{\rm E2/M1}$ extracted
for the bare vertices to the value of $\delta_{\rm E2/M1}$ extracted
for pion dressed vertices has been made in~\ref\dress{ A.M.\ Bernstein,
S.\ Nozawa and M.A.\ Moinester, to be published in Phys.\ Rev.\ {\bf C}
(1993).}, using the model of ref.~\photod.  However, it is clear that
the value of $\delta_{\rm E2/M1}$ extracted from the dressed vertex is
the only quantity with any physical significance; the ratio of bare
vertices will be renormalization scheme dependent in chiral
perturbation theory, and highly model--dependent in any quark or bag
model.  The lattice calculations~\latta, on the other hand, do dress
the vertex to some extent, and do not make an unphysical separation
into local counterterm and loop contributions.  However, by treating
the $\Delta$ and the hyperon resonances as stable particles they do not
include the final state interactions which lead to an imaginary
contribution to the ratio.

There is also the question of $\Theta_2$, which was neglected in our
computation.   Formally its contribution to $\delta_{\rm E2/M1}$ is
subdominant to the contribution from pion loops, which are enhanced by
a chiral log.  However, we should note that our counterterm $\Theta_1$
(in the $\overline{\rm MS}$ subtraction scheme) is approximately three
times larger than we would naively expect~\butlera, and if nature were
to be particularly unkind,  $\Theta_2$ could also be larger than our
naive estimate.  If this is in fact true then we would find
disagreement between our predictions and experimentally determined
values of  $\delta_{\rm E2/M1}$ for the SU(3) allowed transitions.
However, by measuring $\delta_{\rm E2/M1}$ for one of the allowed
transitions the counterterm $\Theta_2$ could be fixed, enabling us to
predict the mixing ratio of the other SU(3) allowed decays.  It is
important to stress that this does not affect in any way our prediction
for the SU(3) forbidden decays, which do not receive contributions from
either counterterm.

The allowed region for each mixing ratio, shown in \allow\ and \supp\ ,
was determined from the coupling constants $F, D, \CC$ and $\CH$, and
associated uncertainties, and from the radiative width of
$\Delta\rightarrow N\gamma$ and its uncertainty.  It is important to
realise that there are other uncertainties in these predictions arising
from terms higher order in the chiral expansion that are not shown in
\allow\ and \supp. An example is the two-loop contribution to the E2
matrix element.  These will generically be $\sim 25\%$ of the lowest
order result in addition to the uncertainties shown in \allow\ and
\supp\ .  One of these uncertainties arising from higher order terms is
exhibited in the choice of the $\omega_\gamma$ used in evaluating the
loop integrals and the formula for the  overall decay rate.  In the
infinite baryon mass limit the photon energy is simply equal to the
mass difference between the initial and final state baryon.
Corrections to this relation, however, occur at order $1/M_B$ and are
hence higher order in the chiral expansion.   In reality, these are
about a 10\% correction but become potentially important when
$\omega_\gamma$ is raised to a high power.

In conclusion, we have used heavy baryon chiral perturbation theory to
compute the leading contribution to the ratio of electric quadrupole to
magnetic dipole radiation, $\delta_{\rm E2/M1}$, in the radiative decay
of baryon resonances.  The magnetic dipole transition is dominated by
the local counterterm and  from nonanalytic $m_s^{1/2}$ contributions
from loop graphs involving  kaons.  On the other hand, the electric
quadrupole transition is dominated by pion loop graphs contributing
terms of order $\log (m^2_\pi / \Lambda_\chi^2 )$ and thus by
calculable long-distance physics.  These long distance effects are
formally larger than the naive estimate of the local counterterm for
the E2 amplitude and so our result is formally the dominant
contribution.  We find that the mixing ratio for $\Delta\rightarrow
N\gamma$ is larger than previously estimated in hadronic models such as
the constituent quark model and also  from model dependent extractions
from pion photoproduction, but is consistent in magnitude with the
latest results from  the lattice.  Since we have used a systematic,
consistent field theoretic approach to the calculation and we have been
able to find the leading, formally dominant contribution, we have some
confidence  in   our result.

We have also shown that an experimental determination of $\delta_{\rm
E2/M1}$ for $\Sigma^{*+}\rightarrow\Sigma^+\gamma$ and
$\Xi^{*-}\rightarrow\Xi^-\gamma$ would provide stringent constraints on
deviations from heavy baryon spin-flavour SU(6).  We feel that these
are very exciting predictions and strongly urge experimentalists to
test them.

\bigskip

MJS and MNB wish to thank the physics department at Duke University
for their kind hospitality while some of this work was performed.
We wish to thank A. Manohar and S. Nozawa for useful
discussions.  MNB acknowledges the support of the Natural Science and
Engineering Research Council  (NSERC) of Canada.  MJS acknowledges
the
support of a Superconducting Supercollider National Fellowship from
the
Texas National Research Laboratory Commission under grant  FCFY9219.
RPS acknowledges the support of DOE grant DE-FG05-90ER40592.

\bigskip
\bigskip
\vfil\eject

\listrefs
\listfigs
\end